\documentclass[aps,prd,twocolumn,superscriptaddress,nofootinbib]{revtex4-2}

\usepackage{latexsym}
\usepackage{amsmath}
\usepackage{amssymb}
\usepackage{amsfonts}
\usepackage{bm}
\usepackage{physics}

\usepackage{color}
\definecolor{purple}{rgb}{0.5,0,0.5}
\definecolor{blue}{rgb}{0.0,0,0.9}
\definecolor{prdblue}{rgb}{0.133,0.118,0.498}
\usepackage[colorlinks=true, pdfstartview=FitV, linkcolor=prdblue, citecolor= prdblue, urlcolor=prdblue]{hyperref}

\usepackage{supertabular} 
\usepackage{placeins}
\usepackage{epsfig}
\usepackage{graphicx}

\begin{document}

%\preprint{}

%\title{Stability assessment of multiquark systems using a Diffusion Monte Carlo method}
\title{A diffusion Monte Carlo calculation of fully heavy pentaquarks}


\author{M.C. Gordillo}
\email[]{cgorbar@upo.es}
\affiliation{Departamento de Sistemas F\'isicos, Qu\'imicos y Naturales, Universidad Pablo de Olavide, E-41013 Sevilla, Spain}
\affiliation{Instituto Carlos I de Física Teórica y Computacional,
Universidad de Granada, E-18071 Granada, Spain.}

\author{J. Segovia}
\email[]{jsegovia@upo.es}
\affiliation{Departamento de Sistemas F\'isicos, Qu\'imicos y Naturales, Universidad Pablo de Olavide, E-41013 Sevilla, Spain}

\author{J. M. Alcaraz-Pelegrina}
\email{fa1alpej@uco.es}
\affiliation{Departamento de F\'{\i}sica.
Universidad de C\'ordoba, Campus de Rabanales, Edif. C2 
E-14071 C\'ordoba, Spain}


\date{\today}

\begin{abstract}
Using a diffusion Monte Carlo algorithm,  we calculated the spectra of all possible $S$-wave fully heavy pentaquarks
% $QQQQ\bar Q$, with either $Q=c$ or $b$ 
within the framework of the quark model.
Our aim was to compare the masses of different spin-color configurations corresponding to the same spatial wavefunction for each quark composition.  In particular,  we computed the masses of the configuration with the maximum number of undistinguishable quarks and those of all the possible baryon+meson splittings compatible with the total number of $b$'s and $c$'s for a given pentaquark.  
What we found was that, in all cases,  compact baryon+meson configurations had lower masses than their more symmetric counterparts.  
Moreover,  those "molecular" associations were unstable (with larger masses) that their infinitely separated non-interacting pieces except int the case of $(ccb)(c\bar{b})$ and $(bbc)(b\bar{c})$ pentaquarks.  In any case,  even for unstable arrangements,  the excess masses are so small ($\sim$20-60 MeV) as to make those compact molecular structures serious candidates to be experimentally detected. 

\end{abstract}

%\pacs{}
%\keywords{}

\maketitle

%%%%%%%%%%%%%%%%%%%%%%%%%%%%%%%%%%%%%%%%%%%%%%%%%%%%%%%%%%%%%%%%%%%%%%%%%%%%%%%%
%%%%%%%%%%%%%%%%%%%%%%%%%%%%%%%%%%%%%%%%%%%%%%%%%%%%%%%%%%%%%%%%%%%%%%%%%%%%%%%%

%\noindent\emph{1.- Introduction}.\,---\, 
\section{Introduction}

In 1964, Murray Gell-Mann~\cite{Gell-Mann:1964ewy} and George Zweig~\cite{Zweig:1964CERN} proposed independently a very successful classification scheme for hadrons based on their valence quarks and antiquarks. This scheme categorizes hadrons into mesons (quark-antiquark pairs) and baryons (three quarks), but opens the door to larger multiquark associations.  That possibility has been experimentally confirmed, in particular,  with the discovery of several quarkonium-like exotic XYZ multiquarks ~\cite{ParticleDataGroup:2022pth,Lebed:2016hpi, Chen:2016qju, Chen:2016spr, Ali:2017jda, Guo:2017jvc, Olsen:2017bmm, Liu:2019zoy, Brambilla:2019esw, Yang:2020atz, Dong:2020hxe, Dong:2021bvy, Chen:2021erj, Cao:2023rhu, Mai:2022eur, Meng:2022ozq, Chen:2022asf, Guo:2022kdi, Ortega:2020tng, Huang:2023jec, Lebed:2023vnd, Zou:2021sha}.  Motivated by the observations of pentaquark $P_c$ states~\cite{LHCb:2015yax, LHCb:2016ztz, LHCb:2019kea} including both light and heavy quarks,  and by the discovery of what it is thought to be a $cc\bar{c} \bar{c}$ structure \cite{LHCb:2020bwg},  in this work we are going 
to consider the possible masses and configurations of all-heavy pentaquarks $QQQQ\bar Q$, with $Q$ either $c$ or $b$.

Previous theoretical works on fully-heavy pentaquarks are relatively scarce.
Most of them consider ensembles including only  $c$ or  $b$ particles together with their antiparticle counterparts
~\cite{Zhang:2020vpz,Yan:2021glh,Wang:2021xao,jorge2022,Gordillo:2023tnz,Gordillo:2024sem}, even though a couple of works study pentaquarks including different $c$-$b$ combinations ~\cite{An:2020jix,an2022}.  To model these systems, as in the case of other quark associations, we need an initial 
approximation to their wavefunctions that take into account the fermionic nature of the constituent quarks.  This is typically done by splitting them into a symmetric part depending only on the interparticle distances, and a spin-color one antisymmetric with respect to the interchange of two {\em identical} quarks.  In both Refs. \onlinecite{An:2020jix,an2022} all quarks of the same type (for instance,  $c$) were considered to be identical, what implied a global function antisymmetric with respect to the exchange of  every (in this case) $c$-$c$ pair.  However, this needs not be so.  For instance,  a $cccc \bar{c}$ can be thought as a monolithic structure of 4 undistinguishable $c$'s and a distinguishable $\bar{c}$ or as an association of a $ccc$ baryon plus a $c \bar{c}$ meson.  In this last case,  the $c$ quark in the meson is no longer identical to any of those in the baryon,  producing a wavefunction different than for a  compact $cccc \bar{c}$ structure.  In the case of multiquarks including only $c$ units, that decrease in the  global symmetry produces more stable structures ~\cite{Gordillo:2024sem}. 

In this work,  we will calculate the mass and structure of all possible $S$-wave all-heavy pentaquarks, \emph{i.e.}, $QQQQ\bar Q$, with either $Q=c$ or $b$.  However,  we will go beyond  Refs. \onlinecite{An:2020jix,an2022} in two ways: first, we will use a diffusion Monte Carlo method (DMC) to solve the many-body Schr\"odinger equations that describe the different multiquark systems.  
This approach accounts for multi-particle correlations in the physical observables, and it is able, in principle, to converge to the real ground state of the system under consideration improving the upper level provided by a purely variational technique ~\cite{Bai:2016int,Gordillo:2020sgc,Gordillo:2021bra, Ma:2022vqf, Gordillo:2022nnj, Ma:2023int, Mutuk:2023oyz, Gordillo:2023tnz, Alcaraz-Pelegrina:2022fsi,men24}.  Second,  and following the
results of Ref.  \onlinecite{Gordillo:2024sem}, 
an additional goal of the present work will be to perform stability assessments of all fully-heavy pentaquarks assuming both structures with the maximum possible symmetry and all possible baryon+meson splittings compatible with the compositions of the clusters under consideration.  

\section{Method}
Fully-heavy multiquark systems including $n$ particles can be described by the following Hamiltonian:
\begin{equation}
H = \sum_{i=1}^n\left(m_i+\frac{\vec{p}^{\,2}_i}{2m_i}\right) - T_{\text{CM}} + \sum_{j>i=1}^n V(\vec{r}_{ij}) \,,
\label{eq:Hamiltonian}
\end{equation}
where $m_{i}$ is the quark mass, $\vec{p}_i$ its momentum and $T_{\text{CM}}$ the center-of-mass kinetic energy. 
$V(\vec{r}_{ij})$ is a quark-quark interacting potential that contains both one-gluon exchange and color confining interactions:
\begin{equation}
V(\vec{r}_{ij}) = V_{\text{OGE}}(\vec{r}_{ij}) + V_{\text{CON}}(\vec{r}_{ij})\,,
\end{equation}
with %The one-gluon exchange potential is given by
\begin{align}
V_{\text{OGE}}(\vec{r}_{ij}) &= \frac{1}{4} \alpha_{s} (\vec{\lambda}_{i}\cdot
\vec{\lambda}_{j}) \Bigg[\frac{1}{r_{ij}} \nonumber \\
&
- \frac{2\pi}{3m_{i}m_{j}} \delta^{(3)}(\vec{r}_{ij}) (\vec{\sigma}_{i}\cdot\vec{\sigma}_{j}) \Bigg] \,,
\label{eq:OGE}
\end{align}
and \cite{Bali:2005fu}
\begin{equation}
V_{\text{CON}}(\vec{r}_{ij}) = (b\, r_{ij} + \Delta) (\vec{\lambda}_{i}\cdot
\vec{\lambda}_{j}) \,,
\label{eq:CON}
\end{equation}
where $\alpha_s$ is the strong coupling constant, while $\vec{\lambda}$ and $\vec{\sigma}$ are,  respectively,  the  Gell-Mann  color and  Pauli spin matrices.  As it is customary ,  $\delta^{(3)}(\vec{r}_{ij})$ is replaced by:
\begin{equation}
\delta^{(3)}(\vec{r}_{ij}) \to \kappa \, \frac{e^{-r_{ij}^2/r_0^2}}{\pi^{3/2}r_{0}^3} \,,
\end{equation}
where $r_0 = A \left(\frac{2m_im_j}{m_i+m_j}\right)^B$. 
Of all the possible expressions  that fit the above prescriptions, 
in this work we are using the AL1 potential proposed by Silvestre-Brac and Semay in Ref.~\cite{Semay:1994ht}, applied previously to similar systems 
\cite{Gordillo:2020sgc, Gordillo:2021bra, Gordillo:2022nnj, Gordillo:2023tnz, Gordillo:2024sem}, and whose parameters are given in Table~\ref{tab:parameters}.  

\begin{table}[!t]
\caption{\label{tab:parameters} Quark model parameters used in this work for the AL1 potential defined in  Refs.~\cite{Semay:1994ht, Silvestre-Brac:1996myf}.}
\begin{ruledtabular}
\begin{tabular}{llc}
Quark masses & $m_c$ (GeV) & 1.836 \\
             & $m_b$ (GeV) & 5.227 \\
\hline
OGE          & $\alpha_s$        & 0.3802 \\
             & $\kappa$          & 3.6711 \\
             & $A$ (GeV)$^{B-1}$ & 1.6553 \\
             & $B$               & 0.2204 \\
\hline
CON          & $b$ (GeV$^2$)  &  0.1653 \\
             & $\Delta$ (GeV) & -0.8321 \\
\end{tabular}
\end{ruledtabular}
\end{table}

As indicated above, we used the diffusion Monte Carlo (DMC) method to solve the Schr\"odinger equation derived from Eq. \ref{eq:Hamiltonian}.  The details of the method applied to ensembles of quarks were described extensively in previous works \cite{Gordillo:2020sgc,men24}.  We only need to say here that the initial approximation to the many-body wave function of each multiquark,  the so-called trial function,  was chosen to be:
\begin{align}
%\Phi_{i,\alpha}(\bm{R}) &\equiv 
\Phi(\vec{r}_1,\ldots,\vec{r}_n;s_1,\ldots,s_n;c_1,\ldots,c_n) \nonumber \\
%&
= \phi (\vec{r}_1,\ldots,\vec{r}_n) %\nonumber \\ 
%&
\times \big[ \chi_s (s_1,\ldots,s_n) \otimes \chi_c (c_1,\ldots,c_n) \big] \,,
\end{align}
in accordance with previous literature \cite{Gordillo:2020sgc, Gordillo:2021bra, Gordillo:2022nnj, Gordillo:2023tnz, Gordillo:2024sem}. Here,  $\vec{r}_j$, $s_j$ and $c_j$ stand for the position, spin and color of the $j$-quark inside the $n$-quark cluster. 
In this work, we consider only $S$-wave multiquarks,  what implies that $\phi$ depends only on the distance between pairs of quarks:
\begin{equation}
\phi (\vec{r}_1,\ldots,\vec{r}_n) = \prod_{j>i=1}^{n} \exp(-a_{ij} r_{ij}) \,.
\label{eq:radialwf}
\end{equation} 
%Other alternatives to the radial part of the trial function are not considered since, in principle, the DMC algorithm is able to correct its possible shortcomings and produce the exact masses of the multiquarks~\cite{Gordillo:2020sgc}. 
with $a_{ij}$'s fixed in accordance to the boundary conditions of the problem.
The spin and color terms, $\chi_s$ and $\chi_c$, of the wave function are linear combinations of the eigenvectors of the spin and color operators:
\begin{equation}
F^2 = \left(\sum_{i=1}^{n} \frac{\lambda_i}{2} \right)^2\,, \quad\quad
S^2 = \left(\sum_{i=1}^{n} \frac{\sigma_i}{2}\right)^2, 
\label{eq:operators}
\end{equation}
with eigenvalues $F^2$=0 (colorless functions) and $S$=1/2, 3/2, or 5/2 depending on the particular pentaquark considered. 
Since quarks are fermions,  the trial function has to be antisymmetric with respect to the interchange of any two {\em undistinguishable} fermions.  Since Eq.~\eqref{eq:radialwf} is symmetric with respect to the exchange of any two identical quarks,  that restriction has to be applied to the spin-color functions obtained as direct products of the eigenvectors of the above operators.  % are obtained by %,  this is done by applying
%applying the %  Eq.~\eqref{eq:radialwf} is symmetric with respect to the exchange of any two identical quarks, we have to produce spin-color combinations which are antisymmetric with respect to those exchanges in order to fulfill Pauli statistics. To do so, we apply the
This means that those spin-color functions have to be eigenvalues of the 
antisymmetry operator:
\begin{equation}
\mathcal{A} = \frac{1}{N} \sum_{{\alpha}=1}^N (-1)^P \mathcal{P_{\alpha}} \,,
\label{eq:antisymope}
\end{equation}
with an eigenvalue equal to one.   
%to each set of spin-color functions with defined values of $F^2$ and $S^2$. 
In Eq.~\eqref{eq:antisymope}, $N$ is the number of possible quark permutations, $P$ is their order, and $\mathcal{P_{\alpha}}$ stands for any of  the matrices that define those permutations. 
%The input of the DMC calculation will be result of multiplying any of the eigenvectors of this operator with eigenvalues equal to one by the radial part of the trial function given in Eq. \ref{eq:radialwf}. 
%%%%%%%%%%%%%%%%%%%%%%%%%%%%%%%%%%%%%%%%%%%%%%%%%%%%%%%%%%%%%%%%%%%%%%%%%%%%%%%%

\begin{table*}[!t]
\caption{\label{tab:results1} Masses, $M$, in MeV, of the $S$-wave $cccc\bar Q$ and $bbbb\bar Q$ pentaquarks, with $Q$ either $c$ or $b$ for different quark configurations. We also provide relevant interquark mean-square radii, in fm$^2$. The subindexes in $\langle r_{ij}^2 \rangle$ represent $i$-quark and $j$-(anti)quark within the $[1234\bar{5}]$-pentaquark.
No structures with four undistinguishable quarks for $J^P=5/2^-$ are considered, since there is no spin-color function antisymmetric with respect the interchanges of any two quarks.In all cases, the error bars correspond to the last figure of the respective values.  } 
\begin{ruledtabular}
\begin{tabular}{lcccccc}
\multicolumn{7}{c}{$cccc\bar c$ pentaquarks} \\[0.5ex]
$J^P$ & Configuration &$M$ & $\langle r_{12}^2 \rangle$ & $\langle r_{14}^2 \rangle$ & $\langle r_{15}^2 \rangle$ & $\langle r_{45}^2 \rangle$ \\
\hline
$\frac{1}{2}^-$ & $(cccc)\bar c$ &$8194$ & $0.30$ & $\cdots$ & $0.31$ & $\cdots$ \\
& $(ccc)(c\bar c)$ &$7923$ &$0.23$ & $0.61$ & $\cdots$ & $0.17$ \\
& Threshold & $7899$ & $\cdots$ & $\cdots$ & $\cdots$ & $\cdots$ \\
%& Mass difference & $24$ \\
\hline
$\frac{3}{2}^-$ & $(cccc)\bar c$ &$8151$ & $0.30$ & $\cdots$ & $0.29$ & $\cdots$ \\
& $(ccc)(c\bar c)$ & $7824$ & $0.23$ & $0.64$ & $\cdots$ & $0.14$ \\
& Threshold & $7803$ & $\cdots$ & $\cdots$ & $\cdots$ & $\cdots$ \\
%& Mass difference & $19$ \\
\hline
$\frac{5}{2}^-$ & $(ccc)(c\bar c)$  &$7899$ & $0.24$ & $\infty$ & $\cdots$ & $0.18$ \\
& Threshold &$7899$ & $\cdots$ & $\cdots$ & $\cdots$ & $\cdots$ \\
%& Mass difference & $0$ \\
\hline
\hline
\multicolumn{7}{c}{$cccc\bar b$ pentaquarks} \\[0.5ex]
$J^P$ & Configuration & $M$ & $\langle r_{12}^2 \rangle$ & $\langle r_{14}^2 \rangle$ & $\langle r_{15}^2 \rangle$ & $\langle r_{45}^2 \rangle$ \\
\hline
$\frac{1}{2}^-$ & $(cccc)\bar b$ & $11437$ & $0.28$ & $\cdots$ & $0.22$ & $\cdots$ \\
& $(ccc)(c\bar b)$ &$11160$ & $0.22$ & $0.50$ & $0.46$ & $0.11$ \\
& Threshold  &$11140$ & $\cdots$ & $\cdots$ & $\cdots$ & $\cdots$ \\
%& Mass difference & $17$ \\
\hline
$\frac{3}{2}^-$ & $(cccc)\bar b$ &$11417$ & $0.28$ & $\cdots$ & $\cdots$ & $0.22$ \\
& $(ccc)(c\bar b)$  &$11111$ & $0.22$ & $0.47$ & $0.44$ & $0.10$ \\
& Threshold &$11090$ & $\cdots$ & $\cdots$ & $\cdots$ & $\cdots$ \\
%& Mass difference & $19$ \\
\hline
$\frac{5}{2}^-$ & $(ccc)(c\bar b)$ & $11140$ & $0.24$ & $\infty$ & $\cdots$ & $0.11$ \\
& Threshold &  $11140$ & $\cdots$ & $\cdots$ & $\cdots$ & $\cdots$ \\
%& Mass difference & $0$ \\
\hline
\hline
\multicolumn{7}{c}{$bbbb\bar b$ pentaquarks} \\[0.5ex]
$J^P$ & Configuration & $M$ & $\langle r_{12}^2 \rangle$ & $\langle r_{14}^2 \rangle$ & $\langle r_{15}^2 \rangle$ & $\langle r_{45}^2 \rangle$ \\
\hline
$\frac{1}{2}^-$ & $(bbbb)\bar b$ & $24211$ & $0.09$ & $\cdots$ & $0.09$ & $\cdots$ \\
& $(bbb)(b\bar b)$ & $23898$ & $0.06$ & $0.14$ & $\cdots$ & $0.04$ \\
& Threshold & $23860$ & $\cdots$ & $\cdots$ & $\cdots$ & $\cdots$ \\
%& Mass difference & $39$ \\
\hline
$\frac{3}{2}^-$ & $(bbbb)\bar b$ & $24192$ & $0.09$ & $\cdots$ & $0.09$ & $\cdots$ \\
& $(bbb)(b\bar b)$ & $23878$ & $0.06$ & $0.13$ & $\cdots$ & $0.04$ \\
& Threshold & $23822$ & $\cdots$ & $\cdots$ & $\cdots$ & $\cdots$ \\
%& Mass difference & $19$ \\
\hline
$\frac{5}{2}^-$ & $(bbb)(b\bar b)$ & $23860$ & $0.07$ & $\infty$ & $\cdots$ & $0.05$ \\
& Threshold & $23860$ & $\cdots$ & $\cdots$ & $\cdots$ & $\cdots$ \\
%& Mass difference & $0$ \\
\hline
\hline
\multicolumn{7}{c}{$bbbb\bar c$ pentaquarks} \\[0.5ex]
$J^P$ & Configuration & $M$ & $\langle r_{12}^2 \rangle$ & $\langle r_{14}^2 \rangle$ & $\langle r_{15}^2 \rangle$ & $\langle r_{45}^2 \rangle$ \\
\hline
$\frac{1}{2}^-$ & $(bbbb)\bar c$ & $21046$ & $0.09$ & $\cdots$ & $0.16$ & $\cdots$ \\
& $(bbb)(b\bar c)$ & $20775$ & $0.06$ & $0.19$ & $0.23$ & $0.09$ \\
& Threshold & $20741$ & $\cdots$ & $\cdots$ & $\cdots$ & $\cdots$ \\
%& Mass difference & $34$ \\
\hline
$\frac{3}{2}^-$ & $(bbbb)\bar c$ & $21018$ & $0.10$ & $\cdots$ & $0.15$ & $\cdots$ \\
& $(bbb)(b\bar c)$ & $20714$ & $0.06$ & $0.20$ & $0.24$ & $0.09$ \\
& Threshold & $20690$ & $\cdots$ & $\cdots$ & $\cdots$ & $\cdots$ \\
%& Mass difference & $24$ \\
\hline
$\frac{5}{2}^-$ & $(bbb)(b\bar c)$ & $20741$ & $0.07$ & $\infty$ & $\cdots$ & $0.1$ \\
& Threshold & $20741$ & $\cdots$ & $\cdots$ & $\cdots$ & $\cdots$ \\
%& Mass difference & $0$ \\
\end{tabular}
\end{ruledtabular}
\end{table*}

\begin{table*}[!t]
\caption{\label{tab:results2} Masses, in MeV, and relevant interquark mean-square radii, in fm$^2$, of  the $S$-wave $cccb\bar Q$ pentaquarks, with $Q$ either $b$ or $c$. 
Indexes and error bars have the same meaning as in Table~\ref{tab:results1}.The thresholds marked with $*$ correspond to $S$= 1/2 and $S$= 0 for the spins of baryons and mesons, respectively.  Otherwise,  they correspond to baryons with $S$ = 3/2 and mesons with the appropriate spin.  Masses of compact structures below those of their respective independent  thresholds are given in bold. }
\begin{ruledtabular}
\begin{tabular}{lcccccc}
\multicolumn{7}{c}{$cccb\bar b$ pentaquarks} \\[0.5ex]
$J^P$ & Configuration & $M$ & $\langle r_{12}^2 \rangle$  & $\langle r_{14}^2 \rangle$ & $\langle r_{34}^2 \rangle$ & $\langle r_{45}^2 \rangle$ \\
\hline
$1/2^-$ & $(ccc)(b\bar b)$ & $14284$ & $0.22$  & $0.35$ & $0.35$ & $0.04$ \\
& Threshold & $14260$ & $\cdots$ & $\cdots$ & $\cdots$ & $\cdots$ \\
& $(ccb)(c\bar b)$ & ${\bf 14206} $ & $0.20$ & $0.19$ & $0.24$ & $0.23$ \\
& Threshold & $14310^*$ & $\cdots$ & $\cdots$ & $\cdots$ & $\cdots$ -\\
\hline
$3/2^-$ & $(ccc)(b\bar b)$ & $14253$ & $0.22$  & $0.36$ & $0.36$ & $0.04$ \\
& Threshold & $14222$ & $\cdots$ &  $\cdots$ & $\cdots$ & $\cdots$ \\
& $(ccb)(c\bar b)$ & ${\bf 14256}$ & $0.20$ & $0.20$ & $0.23$ & $0.22$ \\
& Threshold & $14338$ & $\cdots$ & $\cdots$ & $\cdots$ & $\cdots$ \\
\hline
$5/2^-$ & $(ccc)(b\bar b)$ & $14260$ & $0.24$ & $\infty$ & $\cdots$ & $0.04$ \\
& Threshold & $14260$ & $\cdots$ & $\cdots$ & $\cdots$ & $\cdots$ \\
& $(ccb)(c\bar b)$ & ${\bf 14298} $ & $0.22$ & $0.23$ & $0.27$ & $0.26$ \\
& Threshold & $14388$ & $\cdots$ & $\cdots$ & $\cdots$ & $\cdots$ \\
\hline
\hline
\multicolumn{7}{c}{$cccb\bar c$ pentaquarks} \\[0.5ex]
$J^P$ & Configuration & $M$ & $\langle r_{12}^2 \rangle$ & $\langle r_{14}^2 \rangle$ & $\langle r_{34}^2 \rangle$ & $\langle r_{45}^2 \rangle$ \\
\hline
$1/2^-$ & $(ccc)(b\bar c)$ & $11164$ & $0.22$ & $0.44$ & $0.44$ & $0.11$ \\
& Threshold & $11141$ & $\cdots$ & $\cdots$ & $\cdots$ & $\cdots$ \\
& $(ccb)(c\bar c)$ & $11061$ & $0.20$ & $0.28$ & $0.26$ & $0.20$ \\
& Threshold & $11023^*$ & $\cdots$ & $\cdots$ & $\cdots$ & $\cdots$ \\
\hline
$3/2^-$ & $(ccc)(b\bar c)$ & $11108$ & $0.22$ & $0.45$ & $0.45$ & $0.10$ \\
& Threshold & $11090$ & $\cdots$ & $\cdots$ & $\cdots$ & $\cdots$ \\
& $(ccb)(c\bar c)$ & $11086$ & $0.20$ & $0.31$ & $0.28$ & $0.19$ \\
& Threshold & $11051$ & $\cdots$ & $\cdots$ & $\cdots$ & $\cdots$ \\
\hline
$5/2^-$ & $(ccc)(b\bar c)$ & $11141$ & $0.24$ & $\infty$ & $\cdots$ & $0.11$ \\
& Threshold & $11141$ & $\cdots$ & $\cdots$ & $\cdots$ & $\cdots$ \\
& $(ccb)(c\bar c)$ & $11160$ & $0.21$ & $0.40$ & $0.39$ & $0.24$ \\
& Threshold & $11147$ & $\cdots$ & $\cdots$ & $\cdots$ & $\cdots$ \\
\end{tabular}
\end{ruledtabular}
\end{table*}

\begin{table*}[!t]
\caption{\label{tab:results3} Same as in  Table \ref{tab:results2} for their $bbbc \bar Q$ counterparts. }
\begin{ruledtabular}
\begin{tabular}{lcccccc}
\multicolumn{7}{c}{$bbbc\bar c$ pentaquarks} \\[0.5ex]
$J^P$ & Configuration & $M$ & $\langle r_{12}^2 \rangle$ & $\langle r_{14}^2 \rangle$ & $\langle r_{34}^2 \rangle$ & $\langle r_{45}^2 \rangle$ \\
\hline
$1/2^-$ & $(bbb)(c\bar c)$ & $17523$ & $0.07$ & $0.40$ & $0.40$ & $0.16$ \\
& Threshold & $17499$ & $\cdots$ & $\cdots$ & $\cdots$ & $\cdots$ \\
& $(bbc)(b\bar c)$ & ${\bf 17448}$ & $0.07$ & $0.07$ & $0.21$ & $0.06$ \\
& Threshold & $17507^*$ & $\cdots$ & $\cdots$ & $\cdots$ & $\cdots$ \\
\hline
$3/2^-$ & $(bbb)(c\bar c)$ & $17435$ & $0.07$ & $0.32$ & $0.32$ & $0.13$ \\
& Threshold & $17403$ & $\cdots$ & $\cdots$ & $\cdots$ & $\cdots$ \\
& $(bbc)(b\bar c)$ & ${\bf 17470}$ & $0.07$ & $0.21$ & $0.21$ & $0.19$ \\
& Threshold & $17539$ & $\cdots$ & $\cdots$ & $\cdots$ & $\cdots$ \\
\hline
$5/2^-$ & $(bbb)(c\bar c)$ & $17499$ & $0.07$ & $\infty$ & $\cdots$ & $0.16$ \\
& Threshold & $17499$ & $\cdots$ & $\cdots$ & $\cdots$ & $\cdots$ \\
& $(bbc)(b\bar c)$ & ${\bf 17536}$ & $0.07$ & $0.09$ & $0.28$ & $0.26$ \\
& Threshold & $17590$ & $\cdots$ & $\cdots$ & $\cdots$ & $\cdots$ \\
\hline
\hline
\multicolumn{7}{c}{$bbbc\bar b$ pentaquarks} \\[0.5ex]
$J^P$ & Configuration & $M$ & $\langle r_{12}^2 \rangle$ & $\langle r_{14}^2 \rangle$ & $\langle r_{34}^2 \rangle$ & $\langle r_{45}^2 \rangle$ \\
\hline
$1/2^-$ & $(bbb)(c\bar b)$ & $20774$ & $0.06$ & $0.25$ & $0.25$ & $0.10$ \\
& Threshold & $20741$ & $\cdots$ & $\cdots$ & $\cdots$ & $\cdots$ \\
& $(bbc)(b\bar b)$ & $20699$ & $0.07$ & $0.10$ & $0.15$ & $0.05$ \\
& Threshold & $20639^*$ & $\cdots$ & $\cdots$ & $\cdots$ & $\cdots$ \\
\hline
$3/2^-$ & $(bbb)(c\bar b)$ & $20730$ & $0.07$ & $0.23$ & $0.23$ & $0.09$ \\
& Threshold & $20690$ & $\cdots$ & $\cdots$ & $\cdots$ & $\cdots$ \\
& $(bbc)(b\bar b)$ & $20728$ & $0.07$ & $0.10$ & $0.15$ & $0.05$ \\
& Threshold & $20671$ & $\cdots$ & $\cdots$ & $\cdots$ & $\cdots$ \\
\hline
$5/2^-$ & $(bbb)(c\bar b)$ & $20741$ & $0.07$ & $\infty$ & $\cdots$ & $0.11$ \\
& Threshold & $20741$ & $\cdots$ & $\cdots$ & $\cdots$ & $\cdots$ \\
& $(bbc)(b\bar b)$ & $20722$ & $0.08$ & $0.18$ & $0.24$ & $0.06$ \\
& Threshold & $20709$ & $\cdots$ & $\cdots$ & $\cdots$ & $\cdots$ \\
\end{tabular}
\end{ruledtabular}
\end{table*}

\begin{table*}[!t]
\caption{\label{tab:results4} Same as in the previous Tables, but for the  $S$-wave $ccbb\bar Q$ pentaquarks, with $Q$ either $c$ or $b$.  The thresholds marked with $*$ correspond to $S$=1/2 and $S$=0 for the spins of baryons and mesons, respectively. }
\begin{ruledtabular}
\begin{tabular}{lcccccc}
\multicolumn{7}{c}{$ccbb\bar c$ pentaquarks} \\[0.5ex]
$J^P$ & Configuration & $M$ & $\langle r_{12}^2 \rangle$ & $\langle r_{14}^2 \rangle$ & $\langle r_{15}^2 \rangle$ & $\langle r_{45}^2 \rangle$ \\
\hline
$1/2^-$ & $(ccb)(b\bar c)$ & $14348$ & $0.19$ & $0.20$ & $0.26$ & $0.13$ \\
& Threshold & $14310^*$ & $\cdots$ & $\cdots$ & $\cdots$ & $\cdots$ \\
& $(cbb)(c\bar c)$ & $14248$ & $0.20$ & $0.34$ & $0.26$ & $0.22$ \\
& Threshold & $14220^*$ & $\cdots$ & $\cdots$ & $\cdots$ & $\cdots$ \\
\hline
$3/2^-$ & $(ccb)(b\bar c)$ & $14369$ & $0.19$ & $0.20$ & $0.27$ & $0.14$ \\
& Threshold & $14338$ & $\cdots$ & $\cdots$ & $\cdots$ & $\cdots$ \\
& $(cbb)(c\bar c)$ & $14276$ & $0.19$ & $0.34$ & $0.26$ & $0.22$ \\
& Threshold & $14252$ & $\cdots$ & $\cdots$ & $\cdots$ & $\cdots$ \\
\hline
$5/2^-$ & $(ccb)(b\bar c)$ & $14403$ & $0.20$ & $0.26$ & $0.35$ & $0.17$ \\
& Threshold & $14389$ & $\cdots$ & $\cdots$ & $\cdots$ & $\cdots$ \\
& $(cbb)(c\bar c)$ & $14364$ & $0.24$ & $0.42$ & $0.32$ & $0.28$ \\
& Threshold & $14348$ & $\cdots$ & $\cdots$ & $\cdots$ & $\cdots$ \\
\hline
\hline
\multicolumn{7}{c}{$ccbb\bar b$ pentaquarks} \\[0.5ex]
$J^P$ & Configuration & $M$ & $\langle r_{12}^2 \rangle$ & $\langle r_{14}^2 \rangle$ & $\langle r_{15}^2 \rangle$ & $\langle r_{45}^2 \rangle$ \\
\hline
$1/2^-$ & $(ccb)(b\bar b)$ & $17486$ & $0.18$ & $0.16$ & $0.17$ & $0.07$ \\
& Threshold & $17442^*$ & $\cdots$ & $\cdots$ & $\cdots$ & $\cdots$ \\
& $(cbb)(c\bar b)$ & $17536$ & $0.15$ & $0.25$ & $0.17$ & $0.13$ \\
& Threshold & $17507^*$ & $\cdots$ & $\cdots$ & $\cdots$ & $\cdots$ \\
\hline
$3/2^-$ & $(ccb)(b\bar b)$ & $17520$ & $0.19$ & $0.16$ & $0.17$ & $0.07$ \\
& Threshold & $17470$ & $\cdots$ & $\cdots$ & $\cdots$ & $\cdots$ \\
& $(cbb)(c\bar b)$ & $17571$ & $0.17$ & $0.25$ & $0.16$ & $0.15$ \\
& Threshold & $17539$ & $\cdots$ & $\cdots$ & $\cdots$ & $\cdots$ \\
\hline
$5/2^-$ & $(ccb)(b\bar b)$ & $17532$ & $0.22$ & $0.20$ & $0.19$ & $0.11$ \\
& Threshold & $17508$ & $\cdots$ & $\cdots$ & $\cdots$ & $\cdots$ \\
& $(cbb)(c\bar b)$ & $17603$ & $0.22$ & $0.32$ & $0.21$ & $0.18$ \\
& Threshold & $17590$ & $\cdots$ & $\cdots$ & $\cdots$ & $\cdots$ \\
\end{tabular}
\end{ruledtabular}
\end{table*}

%%%%%%%%%%%%%%%%%%%%%%%%%%%%%%%%%%%%%%%%%%%%%%%%%%%%%%%%%%%%%%%%%%%%%%%%%%%%%%%%
\section{Results}

Let us begin with the $S$-wave $cccc\bar Q$ and $bbbb\bar Q$ pentaquarks, with $Q$ either $c$ or $b$. Their respective quantum numbers, masses and relevant  mean squared distances between quark pairs are shown in Table~\ref{tab:results1}. 
There,  $(QQQQ)\bar Q^{(\prime)}$ represents a configuration with four undistinguishable quarks, while 
a baryon+meson set is denoted by $(QQQ) (Q\bar Q^{(\prime)})$. For both cases, we solved the {\em same} Schr\"odinger equation derived from the Hamiltonian in Eq. \ref{eq:Hamiltonian}, using as the part of 
the trial wavefunction that depends on the particle coordinates the one given in Eq. \ref{eq:radialwf} with the {\em same} set of ten $a_{ij}$'s parameters, all of them different from zero.   The differences between structures were imposed through the symmetry of the spin-color part of the trial.
The configuration $(QQQ) (Q\bar Q^{(\prime)})$ should correspond to a compact ("molecular") structure with quark-(anti)quark distances similar to each other but with three undistinguishable quarks instead of four.  On the other hand, the values listed under "threshold" are the sum of the masses of a baryon and a meson,  obtained by two {\em independent} DMC calculations or taken from Ref. \onlinecite{Gordillo:2020sgc}.  Those baryon and meson masses correspond to states than can be coupled to get associations with the correct quantum numbers.  If several options are possible, we show the one with the lowest mass.  This means that to the masses of baryons made up of identical quarks,  $QQQ$,  and hence with $S$=3/2,  we have to add those of  mesons with $S$=1 for $J^P=1/2^-$ and $J^P=5/2^-$ pentaquarks,  and the one for $S$=0 mesons for the  $J^P=3/2^-$ case.   On the oher hand,  for $QQQ^{(\prime)}$ baryons we checked both $S$=3/2 and $S$=1/2 possibilities coupled to mesons of adequate spin to produce $J^P$'s =1/2,3/2. We also provide the mean squared distances for all the relevant $ij$ pairs of quarks within a $[1234\bar{5}]$ pentaquark.  If all those distances are finite, we have a more or less loosely bound compact object.  On the other hand,  an infinite value signals the splitting of the pentaquark in two subunits irrespectively of the use of the {\em a priori} compact wavefunction given in Eq. \ref{eq:radialwf}.

The main conclusions we can draw from the data in Table~\ref{tab:results1} are: (a) the masses of all  $(QQQQ)\bar Q^{(\prime)}$
structures  are larger than the ones for any baryon+meson arrangement with the came composition; (b) compact baryon+meson structures are loosely bound,  since the distances between identical quarks located in two different subunits are noticeable larger than the ones inside the baryon (or the meson); (c) those compact baryon-meson associations,  when possible, have masses above their corresponding non-interacting thresholds, but the difference is very little,  of at most $\sim$ 60 MeV, but  typically around  $\sim$ 20 MeV; (d) a compact structure for $J^P = 5/2^-$ is impossible; it splits into two separated units even when the initial trial function is compact.   

On the other hand,  $cccb\bar{Q}$ pentaquarks with $Q$ either $c$ or $b$ can be split into $ccc$+ $Q\bar{Q^{(\prime)}}$ or $ccb$+$Q\bar{Q}$ associations.  The masses and the relevant distances to characterize
those clusters are displayed in Table~\ref{tab:results2}.  The first case is pretty similar to the ones already discussed: we have clear molecular baryon+meson compact structures with distances between quarks greater for particles in different subunits and with masses above but very close to their corresponding non-interacting thresholds. 
In this case,  a compact  $J^P = 5/2^-$ structure is also impossible.   However,  $ccb$+$c\bar{b}$ clusters are stable with respect to separations into their constituent units.  Moreover, for that pentaquark composition, we can have a molecular 
$J^P = 5/2^-$ state.  In any case, stable or not,  sets containing a $ccb$ baryon are more compact structures, as it can be seen from their smaller distances between equivalent pairs of quarks.  If we consider their $bbbc\bar{Q}$ counterparts, we can extract similar conclusions by looking at the data in Table~\ref{tab:results3}: only the $bbc$+$b\bar{c}$ sets are stable with respect to division into independent parts.  In the case of the less symmetrical $ccbb\bar{Q}$ clusters, whose related data are displayed in Table~\ref{tab:results4}, we do not have any molecular distribution of lower mass than the corresponding to their non-interacting parts. 

\section{Discussion}

In this work we have proposed several configurations as candidates for bound all-heavy pentaquarks.  The symmetry of those configurations were fixed via the appropriate spin-color functions.  To do so, we started from the direct product of the three colorless eigenvectors of the color Hamiltonian by a variable number of spin functions: 5 for $S$=1/2, 4 for $S$=3/2 and a single one for $S$=5/2.  This means that, for the $S$=1/2 case, we have 15 spin-color possibilities, 12 for $S$=3/2 and 3 when $S$=5/2.  Of those,  if we want to model a $S$=1/2 $cccc\bar{c}$ pentaquark,  we have to keep the single one that is antisymmetric with respect to the interchange between any pair of $c$ quarks.  On the other hand,  for a baryon+meson $(ccc)(c\bar{c})$ configuration we have 3 spin-color wavefuntions with undistinguishable quarks only in the baryon part of the cluster.  All things being equal,  that last situation implies a larger configuration space accessible to the system,  what allows it to explore quark distributions with larger binding energies or,  equivalently,  lower masses. This is what we see for all the pentaquarks in  Table~\ref{tab:results1} with $J^P=1/2^-$.  In the $J^P=3/2^-$ case, the situation is similar with also three spin-color functions adequate to describe the clusters.  

However,  in the $J^P=5/2^-$ case,  the only spin-color function that fulfills the necessary symmetry considerations is the direct product of the one corresponding to a colorless $ccc$ baryon with $S$=3/2 and a colorless $c\bar{c}$ meson with $S$=1.  In that case,  since the pieces are themselves colorless and with definite values of spin,  we found that the smaller mass corresponds to a distribution in which they are infinitely apart.  This explains why,   as it can be seen in Tables~\ref{tab:results1}-\ref{tab:results4},  we do not have $S$=5/2 compact pentaquarks in which the baryon part is of the form $QQQ$,  not even unstable ones.  Conversely,  when  in the splitting of the pentaquark the baryon is of the form $QQQ^{(\prime)}$,  we have two spin-color functions with the right symmetry for the highest value of $J^P$ that are no longer the direct product of colorless subunits.  This allows the existence of compact clusters different from those obtained from a direct product of a baryon and a meson function. However,  those groupings are,  in the majority of stances,  unstable with respect to their splitting into smaller units.  In any case, all the configurations in Table~\ref{tab:results1} and those with a baryonic $QQQ$ form (Tables~\ref{tab:results2}-\ref{tab:results3}) have masses above those of the sum of their constituent parts for any value of $J^P$.  The same can be said of the $ccbb\bar{Q}$ pentaquarks  in Table ~\ref{tab:results4}, unstable with respect to splitting into independent units  by the same factor of $\sim$20-60 MeV. 
 
On the other hand,  in Tables~\ref{tab:results2} and  \ref{tab:results3}, we have several examples  of pentaquarks that are stable with respect to their infinitely separated units.  All of them have the general structure $(QQQ^{(\prime)}) (Q\bar{Q^{(\prime)}})$ 
and a number of possible  spin-color functions larger than 
those of their  $(QQQ)(Q^{(\prime)}\bar{Q^{(\prime)}})$ counterparts: 8 versus 3 for $J^P=1/2^-$,  7 versus 3 for 
$J^P=3/2^-$ and 2 versus 1 for $J^P=5/2^-$.  This suggest that part of the reason would be the large increase in 
the configurational space afforded by a decreasing in the number of undistinguishable quarks.  This would indicate that,  $(QQQ^{(\prime)}) (Q\bar{Q^{(\prime)}})$ compact structures  could be experimentally detected.  Moreover, the small differences between the masses of the other unstable pentaquarks and those of their infinitely separated units,  makes them good candidates to be observed.  

\begin{acknowledgments}
We acknowledge financial support from 
Ministerio Espa\~nol de Ciencia e Innovaci\'on under grant No. PID2022-140440NB-C22;
Junta de Andaluc\'ia under contract Nos. PAIDI FQM-205 and FQM-370 as well as PCI+D+i under the title: ``Tecnolog\'\i as avanzadas para la exploraci\'on del universo y sus componentes" (Code AST22-0001).
The authors acknowledges, too, the use of the computer facilities of C3UPO at the Universidad Pablo de Olavide, de Sevilla.
\end{acknowledgments}

%%%%%%%%%%%%%%%%%%%%%%%%%%%%%%%%%%%%%%%%%%%%%%%%%%%%%%%%%%%%%%%%%%%%%%%%%%%%%%%%

\bibliography{DMC-FullyHeavyPentaquarks}

\end{document}